\documentclass[aps,prd,draft,showpacs]{revtex4}

\begin{document}

\title{Renormalization Group Invariance of the Pole Mass in the Multi-Higgs system}
\author{Chungku Kim}
\date{\today}

\begin{abstract}
We have investigated the renormalization group running of the pole mass in the
multi-Higgs theory in two different types of the gauge fixing conditions. It
turns out that the pole mass when expressed in terms of the Lagrangian
parameters, is invariant under the renormalization group with 
the beta and gamma functions of the symmetric phase.
\end{abstract}
\maketitle

\affiliation {Department of Physics, College of Natural Science, Keimyung
University, Daegu 42601, KOREA}



The pole mass plays an important role in the process where the
characteristic scale is close to the mass shell\cite{Narison} and was shown
to be infrared finite and gauge invariant\cite{Kronfeld}. The physical
quantities such as the pole mass and the beta functions obtained in the
minimal subtraction(MS) scheme was used in determining the Higgs mass bound%
\cite{Kielanowski} or the vacuum stability analysis\cite{stability}. These
beta functions take different forms when calculated in different
renormalization schemes\cite{example} and in most of the cases, they are
obtained in the symmetric phase with the MS scheme\cite{beta} and the pole
masses we are interested in are those in the broken symmetry phase. Hence,
it seems to be necessary to investigate the renormalization group(RG)
behavior of the physical quantities such as the pole mass in the broken
phase under the beta functions in the symmetric phase and it was shown that
if the pole mass was expressed in terms of the Lagrangian parameters in the
symmetric phase it is RG invariant under the beta and gamma functions
obtained in the symmetric phase in the minimal subtractions(MS) scheme in
case of the neutral scalar and the Abelian Higgs model\cite{Kim1}\cite{Kim2}
and also have shown the gauge parameter independence\cite{Kim3}. Recently,
the models with non-minimal Higgs bosons are of interest in the new physics
to solve the question of the origin of the neutrino mass, matter-antimatter
asymmetry in the universe and the nature of dark matters\cite{multi} and in
this paper, we will investigate the RG behavior of the Higgs pole mass in
these multi-Higgs model by generalizing the method that we have used in \cite
{Kim1}\cite{Kim2}.

The pole masses $M_{i}^{2}$ of the multi-Higgs model is determined as \cite
{Quiros} 
\begin{equation}
\left[ \det \Gamma _{pq}(p^{2})\right] _{p^{2}=-M_{i}^{2}}=0,
\end{equation}
where the renormalized inverse propagator $\Gamma _{pq}(p^{2})$ is obtained from the
renormalized effective action in the broken symmetry phase $\Gamma^{BS}$
(See Eqs.(11) and (21)). In order to see the RG behavior of the pole mass of
the multi-Higgs model defined in Eq.(1), we will investigate the RG equation
of the renormalized effective action in two different gauge fixing terms
according to whether the Lagrangian parameters in the symmetric phase
contains parameters that should be identified as a vacuum expectation
values(VEVs) in the broken symmetry phase or not. The former contains the
Feynman gauge and the $\overline{R_{\xi }}$ gauge\cite{R-Bar} and the latter
contains the $R_{\xi }$ gauge\cite{R} and let us consider the former case
first. The renormalized effective action $\Gamma _{S}[\mu ,\left\{ \phi
_{i}\right\} ,\left\{ m_{i}^{2}\right\} ,\left\{ g_{i}\right\} ,\xi ]$ in
the symmetric phase satisfies the RG equation\cite{Ford} due to the
renormalization mass scale $\mu $ independence as 
\begin{equation}
\lbrack D+\gamma _{ij}\phi _{j}\frac{\delta }{\delta \phi _{i}}]\Gamma
_{S}(\mu ,\left\{ \phi _{i}\right\} ,\left\{ m_{i}^{2}\right\} ,\left\{
g_{i}\right\} ,\xi )=0,
\end{equation}
where $\left\{ \phi _{i}\right\} ,$ $\left\{ m_{i}^{2}\right\} $ and $%
\left\{ \text{ }g_{i}\right\} $\ are the classical fields of Higgs, the
masses and the coupling constants of the model and 
\begin{equation}
D\equiv \mu \frac{\partial }{\partial \mu }+\beta _{m_{i}^{2}}\frac{\partial 
}{\partial m_{i}^{2}}+\beta _{g_{i}}\frac{\partial }{\partial g_{i}}+%
\overline{\gamma }\xi \frac{\partial }{\partial \xi },
\end{equation}
with 
\begin{equation}
\mu \frac{\partial m_{i}^{2}}{\partial \mu }=\beta _{m_{i}^{2}},\mu \frac{%
\partial g_{i}}{\partial \mu }=\beta _{g_{i}},\text{ \ }\mu \frac{\partial
\phi _{i}}{\partial \mu }=\gamma _{ij}\phi _{j}\text{ and }\mu \frac{%
\partial \xi }{\partial \mu }=\overline{\gamma }\xi .\text{ }
\end{equation}
It also satisfies the Nielsen identity\cite{Nielsen} which describe the
dependence of the effective action on the gauge parameter $\xi $ generalized
in the multi-Higgs model as 
\begin{equation}
\{\frac{\partial }{\partial \ln \xi }+C_{i}^{\xi }(\phi )\frac{\delta }{%
\delta \phi _{i}}\}\Gamma _{S}(\mu ,\left\{ \phi _{i}\right\} ,\left\{
m_{i}^{2}\right\} ,\left\{ g_{i}\right\} ,\xi )=0.
\end{equation}
In case of the spontaneous symmetry breaking, the mass terms in the
potential term changes sign as $-\frac{1}{2}m_{i}^{2}\phi _{i}^{2}$ and as a
result Higgs field $\phi _{i}$ develop a VEV satisfying 
\begin{equation}
\left[ \frac{\delta \Gamma _{S}(\mu ,\left\{ \phi _{i}\right\} ,\left\{
-m_{i}^{2}\right\} ,\left\{ g_{i}\right\} ,\xi )}{\delta \phi _{l}}\right]
_{\phi _{i}=v}=0,
\end{equation}
which can be solved to obtain the VEV as a function of $\mu ,m^{2}$ and $%
\lambda $ as $v(\mu ,\left\{ m_{i}^{2}\right\} ,\left\{ g_{i}\right\} ,\xi
). $ Although the mass terms in the potential term changes sign as $-\frac{1%
}{2}m_{i}^{2}\phi _{i}^{2},$ we can formally change the sign of the mass
term $\left\{ m_{i}^{2}\right\} $ in Eq.(2) to obtain 
\begin{eqnarray}
0 &=&[\mu \frac{\partial }{\partial \mu }+\beta _{-m_{i}^{2}}\frac{\partial 
}{\partial (-m_{i}^{2})}+\beta _{g_{i}}\frac{\partial }{\partial g_{i}}+%
\overline{\gamma }\xi \frac{\partial }{\partial \xi }+\gamma _{ij}\phi _{j}%
\frac{\delta }{\delta \phi _{i}}]\Gamma _{S}(\mu ,\left\{ \phi _{i}\right\}
,\left\{ -m_{i}^{2}\right\} ,\left\{ g_{i}\right\} ,\xi )  \nonumber \\
&=&[\mu \frac{\partial }{\partial \mu }+\beta _{m_{i}^{2}}\frac{\partial }{%
\partial m_{i}^{2}}+\beta _{g_{i}}\frac{\partial }{\partial g_{i}}+\overline{%
\gamma }\xi \frac{\partial }{\partial \xi }+\gamma _{ij}\phi _{j}\frac{%
\delta }{\delta \phi _{i}}]\Gamma _{S}(\mu ,\left\{ \phi _{i}\right\}
,\left\{ -m_{i}^{2}\right\} ,\left\{ g_{i}\right\} ,\xi ),
\end{eqnarray}
where we have used $\beta _{-m_{i}^{2}}=-\beta _{m_{i}^{2}}.$ This means
that $\Gamma _{S}(\mu ,\left\{ \phi _{i}\right\} ,\left\{ -m_{i}^{2}\right\}
,\left\{ g_{i}\right\} ,\xi )$ satisfies RG equation with the identical RG
functions as those in the symmetric phase $\Gamma _{S}(\mu ,\left\{ \phi
_{i}\right\} ,\left\{ m_{i}^{2}\right\} ,\left\{ g_{i}\right\} ,\xi ).$ Now,
by operating $D$ on Eq.(6) we obtain 
\begin{eqnarray}
0 &=&(Dv_{j})\left[ \frac{\delta ^{2}\Gamma _{S}(\mu ,\left\{ \phi
_{i}\right\} ,\left\{ -m_{i}^{2}\right\} ,\left\{ g_{i}\right\} ,\xi )}{%
\delta \phi _{j}\delta \phi _{l}}\right] _{\phi _{i}=v_{i}}+\left[ D\frac{%
\delta \Gamma _{S}(\mu ,\left\{ \phi _{i}\right\} ,\left\{
-m_{i}^{2}\right\} ,\left\{ g_{i}\right\} ,\xi )}{\delta \phi _{l}}\right]
_{\phi _{i}=v}  \nonumber \\
&=&(Dv_{j})\left[ \frac{\delta ^{2}\Gamma _{S}}{\delta \phi _{j}\delta \phi
_{l}}\right] _{\phi _{i}=v_{i}}+\left[ -\gamma _{li}\frac{\delta \Gamma _{S}%
}{\delta \phi _{i}}-\gamma _{ij}\phi _{j}\frac{\delta ^{2}\Gamma _{S}}{%
\delta \phi _{i}\delta \phi _{l}}\right] _{\phi _{i}=v},
\end{eqnarray}
where we have used Eqs.(6) and (7) and as a result obtain the RG equation
for the VEVs as 
\begin{equation}
Dv_{i}=\gamma _{ij}v_{j}.
\end{equation}
The effective action in the broken symmetry phase $\Gamma ^{BS}$\ is given
by 
\begin{equation}
\Gamma ^{BS}(\mu ,\text{ }\phi _{i},\left\{ m_{i}^{2}\right\} ,\left\{
g_{i}\right\} ,\xi )=\Gamma _{S}(\mu ,\text{ }\phi _{i}+v_{i}(\mu ,\left\{
m_{i}^{2}\right\} ,\left\{ g_{i}\right\} ,\xi ),\left\{ -m_{i}^{2}\right\}
,\left\{ g_{i}\right\} ,\xi ),
\end{equation}
\ and by operating $D$ on $\Gamma ^{BS},$ we can obtain the RG equation for $%
\Gamma ^{BS}$ as 
\begin{eqnarray}
D\Gamma ^{BS} &=&D\Gamma _{S}(\mu ,\text{ }\phi _{i}+v_{i},\left\{
-m_{i}^{2}\right\} ,\left\{ g_{i}\right\} ,\xi )=(Dv_{j})\frac{\partial
\Gamma _{S}(\mu ,\text{ }\phi _{i}+v_{i},\left\{ -m_{i}^{2}\right\} ,\left\{
g_{i}\right\} ,\xi )}{\partial \phi _{i}}+D\Gamma _{S}(\mu ,\text{ }\phi
_{i}+v_{i},\left\{ -m_{i}^{2}\right\} ,\left\{ g_{i}\right\} ,\xi ) 
\nonumber \\
&=&(Dv_{j})\frac{\partial \Gamma _{S}(\mu ,\text{ }\phi _{i}+v_{i},\left\{
-m_{i}^{2}\right\} ,\left\{ g_{i}\right\} ,\xi )}{\partial \phi _{i}}-\gamma
_{ij}(\phi _{j}+v_{j})\frac{\delta \Gamma _{S}(\mu ,\text{ }\phi
_{i}+v_{i},\left\{ -m_{i}^{2}\right\} ,\left\{ g_{i}\right\} ,\xi )}{\delta
\phi _{i}}=-\gamma _{ij}\phi _{j}\frac{\delta \Gamma ^{BS}}{\delta \phi _{i}}%
,
\end{eqnarray}
where we have used Eq.(9) to obtain the last line of above equation. Then
the renormalized inverse propagator $\Gamma _{pq}(p^{2})$ can be obtained from 
\begin{equation}
\Gamma _{pq}(p^{2})=\left[ \frac{\delta ^{2}\Gamma ^{BS}}{\delta \phi
_{p}\delta \phi _{q}}\right] _{\phi =0},
\end{equation}
and by applying $\frac{\delta ^{2}}{\delta \phi _{p}\delta \phi _{q}}$\ to
the RG equation of the renormalized effective action in the broken symmetry
phase $\Gamma ^{BS}$ given in Eq.(10), we can obtain the RG equation for $%
\Gamma _{pq}(p^{2})$ as 
\begin{equation}
D\Gamma _{pq}(p^{2})=\gamma _{pi}\Gamma _{iq}(p^{2})+\gamma _{qi}\Gamma _{ip}(p^{2}).
\end{equation}

Now, let us consider the $R_{\xi }$ gauge fixing where the Lagrangian
parameters in the symmetric phase contain parameters that should be
identified as a VEV in the broken symmetry phase and briefly review the case
of Abelian Higgs model with one Higgs \cite{Kim3}. In $R_{\xi }$ gauge in
the symmetric phase, the gauge fixing term is given by 
\begin{equation}
\frac{1}{2\xi }(\partial _{\mu }A_{\mu }-\xi guG)^{2}.
\end{equation}
This gauge fixing term causes a tadpole divergence which should removed by
an additive renormalization of Higgs field in addition to the usual
multiplicative renormalization\cite{Willey} and a result, the gamma function
for the Higgs field in symmetric phase is given by 
\begin{equation}
\mu \frac{\partial \phi }{\partial \mu }=\text{ }\gamma \phi +\widehat{%
\gamma }u.
\end{equation}
One of the consequence of this process is the difference of the RG behavior
between the Higgs field and the VEV\cite{Sperling}. In the broken symmetry
phase, the Higgs field $H$ develops an VEV $v$ and as a result, the kinetic
mixing term arises between the gauge field $A_{\mu }$ and the Goldstone
field $G.$ By identifying $u$ in Eq.(14) with $v$ in the broken symmetry
phase, the cross term cancels the kinetic mixing term in the broken symmetry
phase.

Now let us consider the general multi-Higgs model containing the Higgs
fields $H_{i}$ and the Goldstone bosons $G_{i}$. In this case, the possible
kinetic mixing terms with the gauge boson $A_{\mu i\text{ }}$will be of the
form $A_{\mu i}v_{j}H_{k}$ or $A_{\mu i}v_{j}G_{k}$ where $v_{j}$ is the VEV
of the Higgs field $H_{j}$\ and in order to remove this we need a gauge
fixing terms $\frac{1}{2\xi }(\partial _{\mu }A_{i\mu }-\xi gu_{j}H_{k})^{2}$
and $\frac{1}{2\xi }(\partial _{\mu }A_{\mu }-\xi gu_{j}H_{k})^{2}$ in the
symmetric phase and then should identify $u_{j}$ with $v_{j}$ in the broken
symmetry phase. Then the renormalized effective action can be written as $%
\Gamma _{S}[\mu ,\left\{ \phi _{i}\right\} ,\left\{ g_{i}\right\} ,\left\{
m_{i}^{2}\right\} ,\xi ,\left\{ u_{i}\right\} ]$ where $\left\{ \phi
_{i}\right\} ,\left\{ g_{i}\right\} $ and $\left\{ m_{i}^{2}\right\} $ are
the classical fields of Higgs, coupling constants and the masses of the
model. Since the renormalized effective action $\Gamma _{S}[\mu ,\left\{
\phi _{i}\right\} ,\left\{ c_{i}\right\} ,\xi ,\left\{ u_{i}\right\} ]$ do
not depend on the renormalization mass scale $\mu ,$ it satisfies the RG
equation\cite{Ford} as \ 
\begin{equation}
\lbrack D+(\gamma _{ij}\phi _{j}+\widehat{\gamma }_{ij}u_{j})\frac{\delta }{%
\delta \phi _{i}}+\widetilde{\gamma _{ij}}u_{j}\frac{\partial }{\partial
u_{i}}]\Gamma _{S}(\mu ,\left\{ \phi _{i}\right\} ,\left\{ m_{i}^{2}\right\}
,\left\{ g_{i}\right\} ,\xi ,\left\{ u_{i}\right\} )=0,
\end{equation}
where the operator $D$ is given in Eq.(3) and 
\begin{equation}
\mu \frac{\partial m_{i}^{2}}{\partial \mu }=\beta _{m_{i}^{2}},\mu \frac{%
\partial g_{i}}{\partial \mu }=\beta _{g_{i}},\text{ }\mu \frac{\partial
\phi _{i}}{\partial \mu }=\gamma _{ij}\phi _{j}+\widehat{\gamma }_{ij}u_{j}%
\text{ },\text{ }\mu \frac{\partial \xi }{\partial \mu }=\overline{\gamma }%
\xi \text{ and }\mu \frac{\partial u_{i}}{\partial \mu }=\text{ }\overline{%
\gamma _{ij}}u_{j}.
\end{equation}
Like the gauge parameter $\xi ,$\ the gauge parameters $\left\{
u_{i}\right\} $ satisfy the Nielsen identity \cite{Bazeia} \ as 
\begin{equation}
\frac{\partial \Gamma _{S}(\mu ,\left\{ \phi _{i}\right\} ,\left\{
m_{i}^{2}\right\} ,\left\{ g_{i}\right\} ,\xi ,\left\{ u_{i}\right\} )}{%
\partial \ln u_{i}}+C_{ik}^{u}(\phi )\frac{\partial \Gamma _{S}(\mu ,\left\{
\phi _{i}\right\} ,\left\{ m_{i}^{2}\right\} ,\left\{ g_{i}\right\} ,\xi
,\left\{ u_{i}\right\} )}{\partial \phi _{k}}=0.
\end{equation}
By combining Eqs.(16) and (18) we can write 
\begin{equation}
\lbrack D+(\gamma _{ij}\phi _{j}+\widehat{\gamma }_{ij}u_{j})\frac{\delta }{%
\delta \phi _{i}}-\widetilde{\gamma _{ij}}\frac{u_{j}}{u_{i}}C_{ik}^{u}(\phi
)\frac{\delta }{\delta \phi _{k}}]\Gamma _{S}(\mu ,\left\{ \phi _{i}\right\}
,\left\{ m_{i}^{2}\right\} ,\left\{ g_{i}\right\} ,\xi ,\left\{
u_{i}\right\} )=0.
\end{equation}
Now, in the broken symmetry phase, the mass terms $\left\{ m_{i}^{2}\right\} 
$ change sign and the renormalized effective action in the broken symmetry
phase $\Gamma ^{BS}$ is obtained not only by substituting $\phi _{i}+v_{i}$
for $\phi _{i}$ as in case of Feynman gauge and the $\overline{R_{\xi }}$
gauge but by identifying $u_{i}$ with $v_{i}$ in order to remove the kinetic
mixing term between the gauge field and the Goldstone field so that 
\begin{equation}
\Gamma ^{BS}=\left[ \Gamma _{S}(\mu ,\left\{ \phi _{i}+v_{i}\right\}
,\left\{ -m_{i}^{2}\right\} ,\left\{ g_{i}\right\} ,\xi ,\left\{
u_{i}\right\} )\right] _{u_{i}=v_{i}}.
\end{equation}
Therefore, there are two sources of the VEV in $\Gamma ^{BS}$ i.e. one from $%
\phi _{i}+v_{i}$ and the other from $u_{i}$ that should identified as $v_{i}$
and hence the VEVs are obtained as 
\begin{eqnarray}
\left[ \frac{\delta \Gamma ^{BS}}{\delta \phi _{i}}\right] _{\phi _{i}=0}
&=&\left[ \frac{\delta \Gamma _{S}(\mu ,\left\{ \phi _{i}+v_{i}\right\}
,\left\{ -m_{i}^{2}\right\} ,\left\{ g_{i}\right\} ,\xi ,\left\{
u_{i}\right\} )}{\delta \phi _{i}}+\frac{\partial \Gamma _{S}(\mu ,\left\{
\phi _{i}+v_{i}\right\} ,\left\{ -m_{i}^{2}\right\} ,\left\{ g_{i}\right\}
,\xi ,\left\{ u_{i}\right\} )}{\partial u_{i}}\right] _{\phi
_{i}=0,u_{i}=v_{i}}  \nonumber \\
&=&\left[ (\delta _{il}-\frac{C_{il}^{u}(\phi _{i}+v_{i})}{u_{i}})\frac{%
\delta \Gamma _{S}(\mu ,\left\{ \phi _{i}+v_{i}\right\} ,\left\{
-m_{i}^{2}\right\} ,\left\{ g_{i}\right\} ,\xi ,\left\{ u_{i}\right\} )}{%
\delta \phi _{l}}\right] _{\phi _{i}=0,u_{i}=v_{i}}=0,
\end{eqnarray}
where we have used Eq.(18). As a result, we can see that although there are
two sources for VEV in $\Gamma _{S},$ the condition to obtain the VEV
becomes 
\begin{equation}
\left[ \frac{\delta \Gamma _{S}(\mu ,\left\{ \phi _{i}\right\} ,\left\{
-m_{i}^{2}\right\} ,\left\{ g_{i}\right\} ,\xi ,\left\{ v_{i}\right\} }{%
\delta \phi _{l}}\right] _{\phi _{i}=v}=0.
\end{equation}
By operating the operator $D$ on Eq.(22) and by using Eqs.(18),(19) and (22)
we obtain 
\begin{eqnarray}
0 &=&(Dv_{j})\left[ \frac{\delta ^{2}\Gamma _{S}}{\delta \phi _{j}\delta
\phi _{l}}+\frac{\delta ^{2}\Gamma _{S}}{\delta \phi _{l}\delta u_{j}}%
\right] _{\phi _{i}=v,u_{i}=v}+\left[ D\frac{\delta \Gamma _{S}}{\delta \phi
_{l}}\right] _{\phi _{i}=v,u_{i}=v}  \nonumber \\
&=&(Dv_{j})\left[ \frac{\delta ^{2}\Gamma _{S}}{\delta \phi _{j}\delta \phi
_{l}}-\frac{1}{u_{j}}\frac{\partial C_{jk}^{u}(\phi )}{\partial \phi _{l}}%
\frac{\delta \Gamma _{S}}{\delta \phi _{k}}-\frac{C_{jk}^{u}(\phi )}{u_{j}}%
\frac{\delta ^{2}\Gamma _{S}}{\delta \phi _{k}\delta \phi _{l}}\right]
_{\phi _{i}=v,u_{i}=v}  \nonumber \\
&&+\left[ -\gamma _{il}\frac{\delta \Gamma _{S}}{\delta \phi _{i}}-(\gamma
_{ij}\phi _{j}+\widehat{\gamma }_{ij}u_{j})\frac{\delta ^{2}\Gamma _{S}}{%
\delta \phi _{i}\delta \phi _{l}}+\widetilde{\gamma _{ij}}\frac{u_{j}}{u_{i}}%
\frac{\delta C_{ik}^{u}(\phi )}{\delta \phi _{l}}\frac{\delta \Gamma _{S}}{%
\delta \phi _{k}}+\widetilde{\gamma _{ij}}\frac{u_{j}}{u_{i}}C_{ik}^{u}(\phi
)\frac{\delta ^{2}\Gamma _{S}}{\delta \phi _{k}\delta \phi _{l}}\right]
_{\phi _{i}=v,u_{i}=v}  \nonumber \\
&=&[(\delta _{jk}-\frac{C_{jk}^{u}(v)}{v_{j}})(Dv_{j})-(\gamma _{jk}+%
\widehat{\gamma }_{jk}-\widetilde{\gamma _{ij}}\frac{C_{ik}^{u}(v)}{u_{i}}%
)v_{j}]\left[ \frac{\delta ^{2}\Gamma _{S}}{\delta \phi _{j}\delta \phi _{l}}%
\right] _{\phi _{i}=v,u_{i}=v},
\end{eqnarray}
from which obtain the RG equation satisfied by the VEV in the multi-Higgs
model as 
\begin{equation}
(\delta _{jk}-\frac{C_{jk}^{u}(v)}{v_{j}})(Dv_{j})=(\gamma _{jk}+\widehat{%
\gamma }_{jk}-\widetilde{\gamma _{ij}}\frac{C_{ik}^{u}(v)}{v_{i}})v_{j}.
\end{equation}
It turns out that in case of $R_{\xi }$ gauge,\ obtaining the RG equation
for the renormalized inverse propagator $\Gamma _{pq}(p^{2})$\ from the RG equation
for the effective action obtained by operating $D$ on $\Gamma ^{BS}$ as we
have done in case of Feynman and the $\overline{R_{\xi }}$ gauge\ is much
complicated compared to obtaining the RG equation directly applying $D$ to
the renormalized inverse propagator defined in Eq.(12) ; 
\begin{eqnarray}
D\Gamma _{pq}(p^{2}) &=&D\left[ \frac{\delta ^{2}\Gamma ^{BS}}{\delta \phi
_{p}\delta \phi _{q}}\right] _{\phi =0}=D\left[ \frac{\delta ^{2}\Gamma
_{S}(\mu ,\phi _{i},\left\{ -m_{i}^{2}\right\} ,\left\{ g_{i}\right\} ,\xi
,\left\{ u_{i}\right\} )}{\delta \phi _{p}\delta \phi _{q}}\right] _{\phi
=v_{i},u_{i}=v}  \nonumber \\
&=&(Dv_{j})\left[ \frac{\delta ^{3}\Gamma _{S}}{\delta \phi _{j}\delta \phi
_{p}\delta \phi _{q}}+\frac{\partial }{\partial u_{j}}\frac{\delta
^{2}\Gamma _{S}}{\delta \phi _{p}\delta \phi _{q}}\right] _{\phi
=v_{i},u_{i}=v}+\left[ \frac{\delta ^{2}}{\delta \phi _{p}\delta \phi _{q}}%
D\Gamma _{S}\right] _{\phi _{i}=v_{i},u_{i}=v}  \nonumber \\
&=&(Dv_{j})(\delta _{jk}-\frac{C_{jk}^{u}(v)}{v_{j}})\left[ \frac{\delta
^{3}\Gamma _{S}}{\delta \phi _{k}\delta \phi _{p}\delta \phi _{q}}\right]
_{\phi =v_{i},u_{i}=v}-\{(\gamma _{ij}+\widehat{\gamma }_{ij})v_{j}-%
\widetilde{\gamma _{ij}}\frac{v_{j}}{v_{i}}C_{ik}^{u}(v)\}\left[ \frac{%
\delta ^{3}\Gamma _{S}}{\delta \phi _{k}\delta \phi _{p}\delta \phi _{q}}%
\right] _{\phi =v_{i},u_{i}=v}  \nonumber \\
&&\text{ }-\gamma _{pi}\Gamma _{iq}(p^{2})-\gamma _{qi}\Gamma _{ip}(p^{2})+%
\widetilde{\gamma _{ij}}\frac{v_{j}}{v_{i}}\left[ \frac{\delta
C_{ik}^{u}(\phi )}{\delta \phi _{q}}\right] _{\phi =v_{i},u_{i}=v}\Gamma
_{pk}(p^{2})  \nonumber \\
&&+\widetilde{\gamma _{ij}}\frac{v_{j}}{v_{i}}\left[ \frac{\delta
C_{ik}^{u}(\phi )}{\delta \phi _{p}}\right] _{\phi =v_{i},u_{i}=v}\Gamma
_{qk}(p^{2})+\widetilde{\gamma _{ij}}\frac{v_{j}}{v_{i}}\left[ \frac{\delta
C_{ik}^{u}(\phi )}{\delta \phi _{p}\delta \phi _{q}}\frac{\delta \Gamma _{S}%
}{\delta \phi _{k}}\right] _{\phi =v_{i},u_{i}=v}  \nonumber \\
&=&-\gamma _{pi}\Gamma _{iq}(p^{2})-\gamma _{qi}\Gamma _{ip}(p^{2})-\widetilde{%
\gamma _{ij}}\frac{v_{j}}{v_{i}}\left[ \frac{\delta C_{ik}^{u}(\phi )}{%
\delta \phi _{q}}\right] _{\phi =v_{i}}\Gamma _{pk}(p^{2})-\widetilde{\gamma
_{ij}}\frac{v_{j}}{v_{i}}\left[ \frac{\delta C_{ik}^{u}(\phi )}{\delta \phi
_{p}}\right] _{\phi =v_{i}}\Gamma _{qk}(p^{2}),
\end{eqnarray}
where we have used Eqs.(19),(22) and (24) to obtain the last line of above
equation. From Eqs.(13) and (25), we can see that the RG equation for the
two point function\ $\Gamma _{pq}(p^{2})$ in the broken symmetry phase can be
written as 
\begin{equation}
D\Gamma _{pq}(p^{2})+\Delta _{pi}\Gamma _{iq}(p^{2})+\Delta _{qi}\Gamma
_{ip}(p^{2})=0,
\end{equation}
where the element of the matrix $\Delta $ is defined as 
\begin{equation}
\Delta _{pi}=\gamma _{pi},
\end{equation}
in Feynman and the $\overline{R_{\xi }}$ gauge and 
\begin{equation}
\Delta _{pi}=\gamma _{pi}+\widetilde{\gamma _{ij}}\frac{v_{j}}{v_{i}}\left[ 
\frac{\delta C_{ik}^{u}(\phi )}{\delta \phi _{q}}\right] _{\phi =v_{i}}
\end{equation}
in $R_{\xi }$ gauge.

\bigskip

Now, in order to investigate the RG behavior of the pole mass defined in
Eq.(1), let us operate $D$\ to this equation to we obtain 
\begin{eqnarray}
0 &=&D\left[ \det \Gamma (p^{2})\right] _{p^{2}=-M_{i}^{2}}=-(DM_{i}^{2})\left[ 
\frac{\delta }{\delta p^{2}}\det \Gamma (p^{2})\right]
_{p^{2}=-M_{i}^{2}}+\left[ D\det \Gamma (p^{2})\right] _{p^{2}=-M_{i}^{2}} 
\nonumber \\
&=&-(DM_{i}^{2})\left[ \frac{\delta }{\delta p^{2}}\det \Gamma (p^{2})\right]
_{p^{2}=-M_{i}^{2}}+\left[ Tr\{\Gamma ^{-1}(p^{2})D\Gamma (p^{2})\}\det \Gamma
(p^{2})\right] _{p^{2}=-M_{i}^{2}}  \nonumber \\
&=&-(DM_{i}^{2})\left[ \frac{\delta }{\delta p^{2}}\det \Gamma (p^{2})\right]
_{p^{2}=-M_{i}^{2}}+\left[ Tr\{adj(\Gamma (p^{2})D\Gamma (p^{2})\}\right]
_{p^{2}=-M_{i}^{2}},
\end{eqnarray}
where $adj(\Gamma (p^{2}))$ is the adjoint of \ $\Gamma (p^{2})$ given as 
\begin{equation}
adj\{\Gamma (p^{2})\}=\Gamma ^{-1}(p^{2})\det \Gamma (p^{2}).
\end{equation}
Note that although $\left[ \det \Gamma (p^{2})\right] _{p^{2}=-M_{i}^{2}}=0$
and hence $\left[ \Gamma ^{-1}(p^{2})\right] _{p^{2}=-M_{i}^{2}}$ does not
exist, $adj\{\Gamma (p^{2})\}$ is finite matrix in this limit. Then, by using
Eq.(26), we\ obtain the second term of the last line of Eq.(29) as 
\begin{equation}
\left[ Tr\{adj(\Gamma (p^{2})D\Gamma (p^{2})\}\right] _{p^{2}=-M_{i}^{2}}=\left[
Tr\{adj(\Gamma (p^{2})(\Delta \Gamma (p^{2})+\Gamma (p^{2})\Delta \}\right]
_{p^{2}=-M_{i}^{2}}=2\left[ Tr(\det \Gamma (p^{2})\Delta )\right]
_{p^{2}=-M_{i}^{2}}=0.
\end{equation}
Finally, by substituting this result back to the last line of Eq.(29), we
can see that 
\begin{equation}
DM_{i}^{2}=0,
\end{equation}
which means that the pole mass of the multi-Higgs model is invariant under
the identical beta and gamma functions obtained in the symmetric phase in
the MS scheme.

In this paper, we have investigated the RG behavior of the pole masses in
the multi-Higgs model and have shown that if the pole masses in the broken
symmetry phase are expressed in terms of the Lagrangian parameters in the
symmetric phase, it is RG invariant under the identical beta and gamma
functions obtained in the symmetric phase in the MS scheme.  Especially in
case of the one-loop order, if we have obtained the pole mass as a function
of the Lagrangian parameters as 
\begin{equation}
M_{i}^{2}=M_{i(0)}^{2}(\left\{ m_{i}^{2}\right\} ,\left\{ g_{i}\right\}
)+M_{i(1)}^{2}(\left\{ m_{i}^{2}\right\} ,\left\{ g_{i}\right\} ,\mu )+\cdot
\cdot \cdot ,
\end{equation}
where $M_{i(0)}^{2}(\left\{ m_{i}^{2}\right\} ,\left\{ g_{i}\right\} )$ and $%
M_{i(1)}^{2}(\left\{ m_{i}^{2}\right\} ,\left\{ g_{i}\right\} ,\mu )$ are
the tree and the one-loop pole mass determined from the condition given in
Eq.(1), then they must satisfy
\begin{equation}
(\beta _{m_{i}^{2}}\frac{\partial }{\partial m_{i}^{2}}+\beta _{g_{i}}\frac{%
\partial }{\partial g_{i}})M_{i(0)}^{2}(\left\{ m_{i}^{2}\right\} ,\left\{
g_{i}\right\} )+\mu \frac{\partial }{\partial \mu }M_{i(1)}^{2}(\left\{
m_{i}^{2}\right\} ,\left\{ g_{i}\right\} ,\mu )=0.
\end{equation}
When there is only one Higgs in the model as in case of Abelian Higgs model
and standard model, $M_{i(0)}^{2}$ becomes $2m^{2}$ and this equation
becomes 
\begin{equation}
2\beta _{m^{2}}+\mu \frac{\partial }{\partial \mu }M_{i(1)}^{2}(\left\{
m_{i}^{2}\right\} ,\left\{ g_{i}\right\} ,\mu )=0.
\end{equation}
Although the one-loop pole mass of Abelian Higgs model given in Ref.[12]
satisfies this equation, some of the previously reported results of pole
mass in standard model do not satisfy this condition and we can expect this
condition to be a criterion for the proper renormalization scheme.


\end{document}